\documentclass[prd,aps,showpacs]{revtex4}
\usepackage{amsmath}
\usepackage{graphicx}
	\usepackage{amssymb}

\begin{document}

\title{A 6 dimensional $(Z_{2})^3$ symmetric model with warped physical  space}
\author{Chetiya Sahabandu, Peter Suranyi, Cenalo Vaz and L.C. Rohana Wijewardhana }
\address{University of Cincinnati, Cincinnati, Ohio, 45221, USA}
\begin{abstract} The Randall-Sundrum model is studied in 6 dimension with AdS$_4$ or dS$_4$ metric in the physical 4 dimensional space.  Two solutions are found, one with induced 5-dimensional gravity terms added to the induced cosmological constant terms. We study the graviton modes in both solutions by transforming the mass eigenvalue equation to a Schrodinger equation with a volcano potential. The spectrum of gravitational excitations depends on the input parameters of the theory, the six dimensional and the effective four-dimensional cosmological constants. The model gives a physically acceptable spectrum if the 4 dimensional cosmological constant is sufficiently small. 
 \end{abstract}
\maketitle

\section{Introduction}Recent intensive study of string theory revived interest in gravitational models with more than 4 space-time dimensions.  Models with extra (compactified or infinite) dimension,  confining particles on three-branes \cite{ADD,ant,randall1,randall2} have been in the forefront of research in particle physics in the past 
few years.  These models have features that, unlike that of traditional string theories, may make the observation of extra dimensions possible in the near future.  One of the most successful models proposed by Randall and Sundrum (RS)  \cite{randall1,randall2} is based on an orbifold solution of the 5 dimensional Einstein equation in AdS space, with a 3-brane fixed at the zero of the fifth coordinate, u (and possibly at another value of $u$, as well), around which the metric has a $Z_2$ symmetry.  This model may provide a solution to  the hierarchy problem of particle physics.  It confines gravitons to the neighborhood of a brane.

The RS model with infinite extra dimensions \cite{randall2} was extended to $D>5$ dimensions by Arkani-Hamed, Dimopoulos, Dvali and Kaloper \cite{arkani}. The model has intersecting ($D-2$)-branes with gravity localized at the intersection.  At long distances, along the intersection, Newton gravity is recovered. With an appropriately chosen curvature Newton's force is modified only at sub-millimiter distances.  The cosmological constants were fine tuned to result in a Minkowski space at the intersection of the (D-2)-branes.

The original RS model is required to satisfy a fine tuning condition if we demand that the 3-brane of our world has Minkowski metric. However, in our world the cosmological constant is probably non-zero and could have had a large value in the past. Thus, it is of interest to investigate the generalization of the RS model when the physical 4-dimensional subspace is not Minkowski (M$_4$), but AdS$_4$ or dS$_4$.  Such a model has been investigated in 5 dimensions~\cite{karch, friedman, xxx, yyy}. In horospheric coordinates, the space is infinite for AdS$_5$, while it is finite for dS$_5$.  For distances much smaller than the 4 dimensional curvature radius and much larger than the five dimensional curvature radius Newtonian gravity is reproduced. 

The aim of the present paper is to extend the $D=6$ case of \cite{karch, friedman, xxx, yyy} to a warped 4 dimensional subspace.   Just like Ref.~\cite{arkani}, we require $Z_2\times Z_2$ orbifold symmetry in the two extra dimensions. In addition, for simplicity,  we also impose $Z_2$ symmetry under the exchange of the two extra coordinates.  In a previous work Chodos and Poppitz~\cite{Chodos} discussed a three brane embedded in a  six dimensional bulk  using a different type of  ansatz.   Their solution also incorporated a nonvanishing cosmological constant on the physical three brane  but the symmetry of their ansatz is AdS$_{4}\times E(2)$. 

Kaloper also found a $D$=6 solution with dS or AdS geometries on the physical 3-brane ~\cite{kaloper}. His solution uses two 4-branes intersecting at an angle different from the right angle.  The model has $Z_2\times Z_2$ symmetry. It is asymmetric in the exchange of the two extra dimensions. Upon the analysis of small fluctuations no confined graviton mode was found in the case of AdS geometry on the physical 3-brane.

In our analysis we have found two new solutions, both with two intersecting perpendicular 4-branes in the action.  They are symmetric to the exchange of the two extra coordinates. These solutions will be discussed in the next section. In Section 3 we investigate small oscillations around the global solutions.  In a  subset of solutions that includes the ground state we are able to transform the mass eigenvalue equation to a Schrodinger form,  with a now well-known volcano potential. We also find the spectrum   of Kaluza-Klein  modes.  In Section 4 we conclude our paper.  Two appendices  contain some details of the calculations.

 \section{Solution of the Einstein equation}
 In 6 dimensions, when the 4 dimensional physical space is Minkowski, a solution can be found for the orbifold problem \cite{arkani},  just like in 5 dimensions, using conformal coordinates.  One cannot however transform the coordinates to a horospheric coordinate system that has been used to solve the the 5 dimensional RS model without fine tuning \cite{xxx,yyy}.
 
Unfortunately, in 6 (or higher) dimensions, when the 4-dimesional space is warped, the metric cannot be brought either to the horospheric, or to the conformal form. Each of these coordinate systems would only contain a single symmetric function of the two extra coordinates, while, as we will soon see, the metric components depend on at least one non-symmetric function of these variables.
In fact, using appropriate coordinate transformations the most general $Z_2\otimes Z_2 \otimes Z_2$ symmetric ansatz for a 6-dimensional metric which is AdS (dS)  in the 4-dimensional subspace (AdS$_4$ or dS$_4$) that, with a further gauge transformation, can be brought to the form
\begin{equation}
ds^2=\gamma_{AB}dx^Adx^B\equiv\Omega^2\left[g_{\mu\nu}dx^\mu dx^\nu+g(u,v)du^2+g(v,u)dv^2\right],
\label{metric6}
\end{equation}
where $g_{\mu\nu}$ is a 4 dimensional metric satisfying the Einstein equation with a four dimensional cosmological constant $\lambda$, which may be positive (dS$_4$ space), zero (Minkowski space), or negative (AdS$_4$ space).  Finally, $\Omega$ is a symmetric function that can be brought to the form of the conformal factor for M$_4$ space~\cite{arkani}, 
\begin{equation}
\Omega=\frac{1}{[1+a(|u|+|v|)]^2}.
\label{omega}
\end{equation}
With this choice, for $M_4$ space, the metric would reduce to the form of Ref.~\cite{arkani}  and $g[u,v]=1$. 

The only unknown function in the metric, $g(u,v)$, has no symmetry property for the exchange of the two coordinates. 
As we will see below, the global solution of the Einstein equation for function $g(u,v)$ contains a single integration constant.  This constant will be fixed later using the Israel junction conditions.

\subsection{Global solution of the Einstein equation}
The nonzero components of the Einstein tensor are ${\cal G}_{\mu\mu}$ with $\mu=t,x,y,z$, ${\cal G}_{uu}$, ${\cal G}_{vv}$, and ${\cal G}_{uv}$. The four components, ${\cal G}_{\mu\mu}$ provide identical second order partial differential equations in variables $u$ and $v$.  It is easy to show that these equations follow, due  to a Jacobi identity, from the rest of the equations. Thus, the bulk equations  ${\cal G}_{uu}=-g_{uu}\Lambda_6$ and ${\cal G}_{uv}=0$ are sufficient to find a unique (up to an integration constant) solution. The equation ${\cal G}_{vv}=-g_{vv}\Lambda_6$ follows from the first equation using the replacement $u\leftrightarrow v$. These equations contain only first order derivatives of the unknown function $g(u,v)$. 

To simplify notations we will use variables $u$ and $v$ rather than $|u|$ and $|v|$.  We may assume for the purpose of finding bulk (6-dimensional) solutions that $u,v>0$. The role of $Z_2\times Z_2$ symmetry is important only when one solves the junction conditions that we will discuss after finding the global solution of the Einstein equation.

The details of the solution are presented in Appendix A.  We just recapitulate the results.  The solution of the system of equations is given in the form 
\begin{equation}
g(u,v)=\frac{1}{S(\sigma)+A(\delta,\sigma)},
\label{soln}
\end{equation}
where we defined $\sigma=u+v$ and $\delta=u-v$. $A(\delta,\sigma)$ is an odd function of $\delta$. $S(\sigma)$ is given by
\begin{equation}
S(\sigma)=\frac{1}{a^2}\left[\frac{(1+a \sigma)^2\lambda}{6 }-\frac{\Lambda_6}{20 }+C (1+a \sigma)^5\right],
\label{sform}
\end{equation}
where $C$ is a yet undetermined integration constant. 

The Taylor series of $A(\delta,\sigma)$ in $\delta$ is 
\begin{equation}
A(\delta,\sigma)=\sum_{k=0}^\infty\delta^{2k+1}\alpha_k(\sigma),
\label{aseries-main}
\end{equation}
where
\begin{equation}
\alpha_0(\sigma)=S'(\sigma).
\label{a0}
\end{equation}
and $\alpha_k$ satisfies the recursion relation
\begin{equation}
\alpha_k(\sigma)=-\frac{1}{S(\sigma)(2k+1)}\sum_{m=0}^{k-1}\alpha_m'(\sigma)\alpha_{k-1-m}(\sigma) .
\label{recursion-main}
\end{equation}

The recursion relation (\ref{recursion-main}) can be used to generate $\alpha_k$ in arbitrary order. The first few $\alpha_k$ are listed in Appendix A. 

 Note now that $g(u,v)$ is unique, except for the choice of the scale $a$ and the integration constant $C$. One of these constants can be fixed by setting the scale of variables $u$ and $v$.  The other will be fixed by the junction conditions to be discussed later. 
 
 As usual, the scale is set by demanding that the metric is Minkowski at $u=v=0$. As $A(0,0)=0$ we find the following value for the conformal scale from (\ref{sform})
 \begin{equation}
a=\sqrt{\frac{\lambda}{6}-\frac{\Lambda_{6}}{20}+C}.
\label{getlambda}
\end{equation}
 Note that a solution may exist for both positive and negative cosmological constants, as long as $\frac{\lambda}{6 }-\frac{\Lambda_6}{20 }+C>0$.

\subsection{Israel junction conditions}
Due to the requirement of $Z_2\times Z_2$ symmetry the derivatives of the components of the metric tensor are not continuous at $u=0$ and $v=0$.  This discontinuity will generate junction terms in the Einstein tensor, localized on the $u=0$ and $v=0$ planes.  These junction terms are:
\begin{eqnarray}
\Delta G_{\mu\mu}^{(6)}&=&-g_{\mu\mu}\delta(v)\left[\frac{8 a (1+a u)}{g(0,u)}-\frac{g_{,v}(u,v)|_{v=0}(1+a u)^2}{g(0,u)g(u,0)}\right]+(u\leftrightarrow v),\nonumber\\
\Delta G_{uu}^{(6)}&=&-\delta(v)g_{uu}\frac{8a(1+au)}{g(0,u)}\nonumber\\
\Delta G_{vv}^{(6)}&=&-\delta(u)g_{uu}\frac{8a(1+av)}{g(0,v)}
  \label{junction6}
\end{eqnarray}
To satisfy the Einstein equations at $u=0$ and at $v=0$ two 4-brane contributions are required to cancel (\ref{junction6}). Brane contributions come from the variation of  $u=0$ and $v=0$ 4-brane terms such as a tension term (i.e. a five dimensional cosmological constant term) and possibly of  dynamical terms. As we will see later the junction conditions will impose three $u$ and $v$ independent constraints on the constants of the theory.  One of these can be used to fix the integration constant $C$. Then the brane contributions must in general contain two additional constants.  One of these is the cosmological constant, $\Lambda_5$.  We choose the other constant as the 5-dimensional gravitational constant, $G_5$. Thus, we propose to add the following terms to the action:
\begin{equation}
S_5=\frac{1}{8\pi G_5}\left[\int d^4xdu \sqrt{-g}R^{(5)}+ \Lambda_5\int d^4xdu \sqrt{-g}+(u\leftrightarrow v)\right],
\label{action5}
\end{equation}
where $R^{(5)}$ is the 5 dimensional Ricci scalar constructed from the induced metric on the $v=0$ or $u=0$ brane.  It is known that adding a four dimensional gravity term to the five dimensional action changes predictions  for post-Newtonian effects, contradicting observations~\cite{DGP}, \cite{vDVZ,zachar}. The question, whether a similar problem arises about our solution with induced 5-dimensional gravity will be taken up in a future publication.  

The variation of (\ref{action5}) gives
\begin{eqnarray}
{\cal G}_{\mu\mu}^{(5)}+g_{\mu\mu}\Lambda_5&=&g_{\mu\mu}\left[\Lambda_5 -\lambda(1+au)^2+\frac{6a^2}{g(u,0)}+\frac{3a}{2}(1+au)\frac{g_{,u}(u,0)}{g^2(u,0)}\right],\nonumber\\
{\cal  G}_{uu}^{(5)}+g_{uu}\Lambda_5&=&g_{uu}\left[\Lambda_5 -2\lambda(1+au)^2+\frac{6a^2}{g(u,0)}\right].
\label{vzerobrane}
\end{eqnarray}
There are similar contributions on the $u=0$ brane, obtained from (\ref{vzerobrane}) by the exchange ($u\leftrightarrow v$). One of the advantages of maintaining the $u\leftrightarrow v$ symmetry is that one needs to deal with two junction conditions rather than four.

It is easy to see that when we substitute our solution $g(u,v)$ into (\ref{junction6}) and (\ref{vzerobrane}) multiplied by $\delta(v)$ the sum of these contributions does not vanish at all values of $u$.  This alone does not mean that the 6-dimensional junction term cannot be canceled by the brane terms of (\ref{action5}).  The gauge of the junction contributions and of the brane contributions may be misaligned.  In the next section we will how that this is indeed the case.

Another way to satisfy the junction conditions would be to add further invariants to the action, such as Lovelock terms, but one would need an infinite number of such terms in the Lagrangian to have a chance to satisfy infinitely many constraints imposed by the junction conditions.

\subsection{Coordinate transformation}

As the six dimensional action is gauge invariant, we can apply a gauge transformation to its contribution to the junction condition.  We require that the gauge transformations satisfy the following conditions:
\begin{enumerate}
\item The $(u\leftrightarrow v)$ exchange symmetry is maintained.  By this requirement we reduce the number of junction equations to be satisfied from four to two.
\item We require that the transformed coordinates are such that $u'$ and $v'$ vanish when $u$ and $v$ vanish, respecitively.  This constraint is required by the junction equations. It implies that the position of the branes is unchanged by the gauge transformation.
\item We will maintain the scale of the coordinates at $u=v=0,$ by requiring that the components of the gauge transformed metric tensor $\tilde g_{uu}$ and $\tilde g_{vv}$ also tend to one when $u,v\to0$.
\end{enumerate}
These requirements lead to the following form of coordinate transformations
\begin{eqnarray}
u&\to &u'=h(u,v)\nonumber\\
v&\to&v'=h(v,u),
\label{transform}
\end{eqnarray}
where $h(u,v)$ is a $C^2$ function, $h(0,v)=0$, $h_{,u}(u,v)|_{u=v=0}=1.$

We will show in Appendix B. that the two junction conditions can be satisfied by an appropriate choice of $h(u,v)$.  While that result is important, the physical consequences of our solution can be explored without the detailed knowledge of the form of the function $h(u,v)$ provided $h(0,0)=0$ and  $h_{,u}(u,v)|_{u=v=0}=1$.  

When one writes down the junction conditions one must take care of the fact that the action is varied with respect to $g_{AB}$, rather than $\tilde g_{AB}$. In other words, certain linear combinations of (\ref{junction6}) enter the junction conditions.  Denoting these appropriate linear combinations by $\Delta R^{(6)}_{AB}$ the junction conditions at $v=0$ are
\begin{equation}
\frac{\sqrt{-\tilde g}}{G_6}\Delta_v R^{(6)}_{AB}+\frac{\sqrt{- g}}{G_5} R^{(5)}_{AB}\delta(v)=0.
\label{junction3}
\end{equation}
 One obtains two independent junction conditions, one for choosing $A=B=\mu$ and one for $A=B=u$ (see Appendix B). The junction contributions are functionals of two independent single variable functions, $f_1(u)=h(u,0)$ and $f_2(u)=h_{,v}(u,v)|_{v=0}$, in an algebraic manner. The only constraints on these functions are $f_1(0)=0$ and $f_1'(0)=f_2(0)=1$.  We must satisfy these three independent constraints. Once these constraints are satisfied one can always find two functions, $f_{i}(u)$, which solve the two junction conditions.  The function $f_2(u)$ enters the junction conditions in a very simple manner: its third power is a multiplier of the 6-dimensional junction contributions.   Taking the two junction conditions at $u=0$ provides two constraints on the constants of the model. Taking the ratio of the two junction conditions eliminates $f_2(u)$ from the resulting condition.  Taking a derivative of the resulting equation with respect to $u$, at $u=0$, provides a third constraint. The function $h(u,v)$ enters each of the three constraints only through the combinations  $f_1(0)=0$, $f_1'(0)$ and $f_2(0)$, all fixed by our constraints on the gauge transformation.  All other derivatives of the junction conditions taken at $u=0$ contain unconstrained derivatives of $h(u,v)$ that are determined by these very equations. 

It is clear now why, in general, we need the two terms in (\ref{action5}) (cosmological constant and gravity).  Recall that besides the coupling constants $G_5$ and $\Lambda_5$ we still have one undetermined integration constant, $C$ , (or, alternatively the conformal scale, $a$, which is related to $C$ through (\ref{getlambda}). The three equations obtained from the junction conditions are just sufficient to fix these three constants. In fact, one could replace the 5-dimensional scalar curvature term of the Lagrangian by another gauge invariant term, such as a Lovelock term, but the scalar curvature term is the simplest choice. 

The two junction conditions taken at $u=0$ provide the following equations 
\begin{eqnarray}
0&=&-\frac{8a}{G_6}-\frac{1}{G_5}\left(9a^2-\frac{1}{2}\lambda -\Lambda_5+\frac{3}{4}\Lambda_6\right)\label{uno}\\
0&=&-\frac{8a}{G_6}-\frac{1}{G_5}\left(2\lambda-6a^2-\Lambda_5\right),
\label{due}
\end{eqnarray}
where we used the expression, obtained from our solution,
\begin{equation}
a g_{,u}(u,0)|_{u=0}=\lambda-10a^2-\frac{1}{2}\Lambda_6.
\end{equation}

The set of equations (\ref{uno}) and (\ref{due}) have two solutions. 
\begin{enumerate}
\item Subtracting the two equations from each other implies
\begin{equation}
a^2=\frac{\lambda}{6}-\frac{\Lambda_6}{20},
\label{ag}
\end{equation}
or in other words, the integration constant $C=0$. Then we also have
\begin{equation}
\Lambda_5=\frac{8aG_5}{G_6}+\lambda+\frac{3}{10}\Lambda_6.
\end{equation}
The third constraint equation, to be discussed below will also determine the 5-dimensional gravitational constant, $G_5$.
\item
Note that if we take the limit $G_5, \Lambda_{5}\to\infty$ (no scalar curvature term on the brane) such that $\tilde \Lambda_{5}=\Lambda_{5}/8\pi G_{5}$ is finite then  (\ref{uno}) and (\ref{due}) become identical, providing an expression for $\tilde\Lambda_5$
\begin{equation}
\tilde\Lambda_5=\frac{a}{\pi G_6}.
\label{lambda5}
\end{equation}
The third constraint equations will fix the yet undetermined $a$ (or, alternatively, fix $C$).
\end{enumerate}

Now for solution 1.  the mere presence of the dynamical term on the $v=0$ and $u=0$ branes, combined with the fact that the metric tensors depend on $|u|$ and $|v|$, respectively, requires the existence of a non-vanishing tension on the $u=v=0$ 3-branes, just like in the 5 dimensional Randall-Sundrum solution.  We denote the tension by $\Lambda_4$. We obtain
\begin{equation}
\Lambda_4=-6\lambda.
\end{equation}
If $\lambda=0$ (non-warped space) the contribution on the  3-brane vanishes and the solution reduces to that of \cite{arkani}.

Let us consider now the third and final constraint equation. As we indicated earlier it is obtained from the derivative of the ratio of the two junction conditions, eliminating $h_{,v}(v,u)|_{v=0}$ from the equation.  To reduce the obtained equations to a constraint on our constants we need to calculate derivatives of $g(u,v)$ at $u=v=0$. These can be determined either using the explicit form of $g(u,v)$ from (\ref{soln}) or directly from taking derivatives  of (\ref{uno}) and (\ref{due}) and setting $u=v=0$.  To simplify matters we write down the the third constraint separately for the first and second solution. For the first solution ($G_5=$finite) we obtain, using (\ref{ag}) 
\begin{equation}
1700\Lambda_5^2-40\Lambda_5(80+27\Lambda_6)+3(500\lambda^2+340\lambda\Lambda_6+57\Lambda_6^2)=0
\label{third}
\end{equation}
For the second solution ($G_5=\infty$), after substituting (\ref{lambda5}), we obtain for the third constraint
\begin{equation}
100a^2-11\lambda+5\Lambda_6=0
\label{third2}
\end{equation}
These equations can be used to calculate the rest of the constants.
All the constants of the theory are determined by the parameters of the 6 dimensional world, $G_6$ and $\Lambda_6$ and of the physical cosmological constant, $\lambda$. The value of parameters found for the two solutions are given in Table 1, where we use the notations $r=-\lambda /6 a^2$ and $w=1+a \sigma.$ 
\begin{table}
\caption{\label{label} Summary of parameters characterizing the two solutions.}
\begin{center}
\begin{tabular}{lcccccc}\hline
{\rm Solution} & $a$ & $C$ & $G_5$ & $\Lambda_5$ ($\tilde\Lambda_{5}$)&$\Lambda_4$&$S(\sigma)$\\ \hline
{\rm Solution I }&$\pm\sqrt{\frac{\lambda}{6}-\frac{\Lambda_6}{20}} $&0&$-\frac{3a G_6}{34}$&$- 10\lambda-\frac{19}{5}\Lambda_6 $&$-6\lambda$&$-rw^2+1+r$\\ 
{\rm Solution II }&$\pm\sqrt{\frac{11\lambda}{100}-\frac{\Lambda_6}{20}}$ & $-\frac{17\lambda}{300} $& $\infty$&   $ \frac{a}{\pi G_6}$&0&$1+ r\left(\frac{33+17w^{5}}{50}-w^{2}\right)$\\  \hline\end{tabular}
\end{center}
\end{table}

We can see from Table 1 that in fact, taking into account the two possible signs for $a$, we have four independent solutions, rather than two. Note that at a special choice of the parameter $r$, $r=-38/41$ the combination $50\lambda+19\Lambda_6=0$ and $\Lambda_5=0$.  Then the geometry on the intersecting branes is asymptotically Minkowski.

\subsection{The range of variables $u$ and $v$}
The range of variables $u$ and $v$ plays an important role in the investigation of small deformations of the above solutions. The linearized deformations determine the spectrum of gravitational excitations and  the nature of the gravitational force among massive objects.

 The form (\ref{soln}) of the solution implies that zeros of  $S(\sigma)\pm A(\delta,\sigma)$ (for both signs) constitute boundaries in the $(u,v)$ space.  Suppose the equation of such a boundary is on the curve $\delta=\delta(\sigma)$. Then $\delta(\sigma)$ satisfies
\begin{equation}
[g(u,v)]^{-1}=S(\sigma)+A(\delta(\sigma),\sigma)=0.
\label{curve-equation}
\end{equation}
  This equation can easily be solved.  Consider that whenever (\ref{curve-equation}) is satisfied, unless $\delta=0$, $[g(v,u)]^{-1}\not=0$. Then the equation ${\cal G}_{uv}=0$, (\ref{two}) of Appendix A implies that $g_{,v}(u,v)$ must also vanish.  Then, differentiating (\ref{curve-equation}) with respect to $\sigma$ and comparing it with the substitution of  $\delta\to\delta(\sigma)$ into the equation $g_{,v}(u,v)=0$ we obtain $\delta'(\sigma)=-1$. Integrating this equation we obtain $\delta(\sigma)=-\sigma+2 c$, where $ c$ is a constant. Owing to the definition of $\delta$ and $\sigma$, the boundary of the domain is determined by a constant value of $u$, $u=c$.  In other words, $(g[c,v])^{-1}=0$ for all admissible values of $v$. This constant can be determined by taking the limit $v\to c$.
    At this point $\delta=A(\delta,\sigma)=0$ and $\sigma=2c$. In other words, the equation 
\begin{equation}
S(2c)=0
\label{roots}
\end{equation}
provides the corresponding value of $c$.

Notice now that when $g_{uu}= g[u,v]$ has a pole at $u=c$ ~ $g_{vv}= g[v,u]$ also has a pole at $v=c$. Then, whenever (\ref{roots}) has a positive solution the range of variables $u$ and $v$ is  the square $|u|,|v|<c$.

Let us examine now the solution of (\ref{roots}) for our four metric solutions. When $a>0$ the boundary $r=0$ separates domains with or without solutions.  However, when $a<0$ the corresponding boundary value is $r=r_0$, where  $r_0=-1$ for Solution I and $r_0=-50/33$ for Solution II. Owing to the identites $r=-6 a^2 \lambda$ and $1-r/r_0=a^2 20\Lambda_6$, these boundaries separate domains with different curvature: $r>0$ corresponds to AdS$_4$ and $r<0$ to dS$_4$ while $r>r_0$ corresponds to AdS$_6$ while $r<r_0$ corresponds to dS$_6$. Thus, the boundaries are:
\begin{enumerate}

\item Solution I, $a>0$

The boundary, $c$, satisfies
\begin{equation}
-r(1+2ca)^2+1+r=0
\label{range1}
\end{equation}
where $r=-\lambda/6 a^2>0$.  Then for $r>0$ (AdS$_4$) the range of variables $u$ and $v$ is
\begin{equation}
|u|,|v|<\frac{1}{2a}\left(\sqrt{\frac{r+1}{r}}-1\right).
\end{equation}. 

For $r<0$ (dS$_4$) the range of $|u|$ and of $|v|$ is infinite.

\item Solution II, $a>0$

(\ref{roots}) has the form
\begin{equation}
-r(1+2ca)^2+1+\frac{33}{50}r+\frac{17}{50}r (1+2ca)^5=0
\label{range2}
\end{equation}
No analytic solution of this fifth order equation exists but at small positive values of $r$ the range of $u$ and $v$ is infinite.  Increasing $r$ we obtain a critical value, above which the range becomes finite, because (\ref{range2}) has a positive solution. This critical value is obtained from the coincidence of the zero (\ref{range2}) with the zero of the derivative $S(\sigma)$. This happens at 
\begin{equation}
r_{\rm min}=\frac{1}{w_{\rm min}^2-\frac{33}{50}-\frac{17}{80}w_{\rm min}^5}\simeq 115.5,
\end{equation}
where 
\begin{equation}
w_{\rm min}=\left(\frac{20}{17}\right)^{1/3}=1+2ac_{\rm min}.
\end{equation}
If $r$ increases, starting from  $r_{\rm min}$ than the boundary value of $|u|$ and $|v|$ decreases from $c_{\rm min}(r_{\rm min})=(1/2a)[(20/17)^{1/3}-1]=\frac{1}{a}0.2095$ to $c_{\rm min}(\infty)=\frac{1}{a}0.0538$ .
For negative values of $r$ (\ref{range2}) always has a positive root, so the range of variables is finite. 
\item Solution I, $a<0$

The largest possible physical range of $w=1+a \sigma$ is $0<w<1$.  There is a solution of the equation $S(\sigma)=0$ only if $r<-1$. This corresponds to dS$_6$ space. 
The solution of the $S(\sigma)=0$ equation provides the following bounds for 
\begin{equation}
|u|,|v|<\frac{1}{2|a|}\left(1-\sqrt{\frac{|r|-1}{|r|}}\right)
\end{equation}
For $r>-1$ the range of $u$ and $v$ is bounded only by the condition $|u|+|v|<1/|a|$ (i.e. finiteness of the conformal factor).

\item Solution II, $a<0$

At $w=1$ $S=1$. At $w=0$ $S=1+33 r/50$. Therefore, if $r>r_0=-50/33$ (AdS$_6$) there is no solution of the $S(\sigma)=0$ equation (it is easy to see that there is no minimum of $S(\sigma)$ lower than 0). However, there is  a solution of the equation $S(\sigma)=0$ for $r<-50/33$ because $S(w=0)<0$.  So $w\geq w_{\rm min}>0$.These values of $r$ correspond to dS$_6$ space ($\Lambda_6>0$).  Just like in the case $a>0$ there is no analytic solution for $w<w_{\rm min}$.
\end{enumerate}
\section{Small oscillations}

Suppose the metric solutions we previously found, denoted ${}^0g_{ab}$, are modified by a small perturbation, such as 
\begin{equation}
g_{ab}(\epsilon)={}^0g_{ab}+h_{ab}.
\end{equation}
As now it is standard in the literature, we will impose the TT (traceless, transverse) gauge conditions on $h_{ab}$. In addition we will impose axial gauge conditions, $h_{au}=h_{av}=0$. Thus, the nonzero components of the TT tensor $h$ are $h_{\mu\nu}$, where $\mu,\nu=t,x,y,z.$ This gauge is also called the Randall-Sundrum gauge.  The advantage of this gauge (in contrast to the TT conditions applied to the `complete oscillations', $\Omega^2 h_{ab}$), as we will see below, is that for the graviton mode $h_{\mu\nu}$ is independent of $u$ and $v$. Then, at least for $a>0$ the conformal factor $\Omega^2$ insures localization.  In general, gravitational excitations are dependent on these variables, as well. They may or may not be localized to the intersection of the branes.  In what follows, we will drop the superscript 0 from the background metric, ${}^0g_{ab}$.

 The components of $h$ satisfy the following wave equation
\begin{equation}
\frac{1}{2}\nabla^A\nabla_A h+\frac{1}{5}\Lambda h=0,
\label{wave}
\end{equation}
where $A=t,x,...,u,v$ and $\nabla^A$ is a covariant derivative in the background metric, $\gamma_{AB}$. 
Equation (\ref{wave}) is separable. Multiplying by the conformal factor we obtain
\begin{equation}
\frac{g^{\mu\nu}}{2}\tilde\nabla_\mu\tilde\nabla_\nu h+\frac{1}{2}\nabla^{u_i}\nabla_{u_i} h+\frac{1}{5}\Lambda h=0,
\label{wave2}
\end{equation}
where the covariant derivatives $\tilde\nabla_\mu$ are taken in the four dimensional metric $g_{\mu\nu}$ and $u_i$ is a shorthand notation for $u$ and $v$. 

Now the first term of (\ref{wave2}) contains only derivatives and metric components dependent on $t,x,y,$ and $z$, while the rest of the terms depend on $u$ and $v$ only.  Using a factorized ansatz for the wave function and introducing the gravitational mass $m$ we obtain the following two equations (Here we assume a specific form for the 4 dimensional metric: Minkowski with a  warp factor $g_{yy}=g_{xx}=-g_{tt}=\omega^2=e^{-2 b z}$ for AdS$_4$ and $g_{ii}=\omega^2=e^{-2 b t}$, where $i$ denotes a spatial coordinate, for dS$_4$)
\begin{eqnarray}
m^2 h&=&g^{\mu\nu}\partial_\mu\partial_\nu h\pm\frac{7}{2}b\,\partial_\xi h-\frac{5}{3}\lambda h,\label{wave3}\\
m^2 h&=&\left[\frac{\lambda\Omega^{-1}}{4a}-\frac{\Lambda_6\Omega}{8a}+\frac{5a\Omega}{4g(u,v)}+\frac{3a\Omega}{4g(v,u)}-\frac{\partial_vg(u,v)}{g(u,v)}\left(\frac{1}{g(u,v)}+\frac{1}{g(v,u)}\right)\right]\partial_uh\nonumber\\&+&\frac{\partial_u^2h}{g(u,v)}+(u\leftrightarrow v).\label{wave4}
\end{eqnarray}
where the upper sign corresponds to dS$_4$ and the lower one to AdS$_4$. We use $\xi=z$ or $t$ and $\lambda=\pm 3 b^2$.  

The solution of (\ref{wave3}) is straightforward.  It describes wave oscillations
\begin{equation}
h_1\sim e^{i(k_x x+k_yy+k_zz-k_tt)},
\label{soln2}
\end{equation}
where $k_t=\sqrt{k_x^2+k_y^2+k_z^2+m^2}$, provided the coordinates are much smaller than the curvature radius, $1/b$, otherwise the wave is modified, as we will discuss it below.  In particular, the $m=0$ mode always exists since $h_2$=const. is a solution of (\ref{wave4}).  It describes a graviton, as long as the corresponding wave function is normalizable.  In what follows, we will examine the solutions of (\ref{wave4}) and their normalization.  

The exact solution of (\ref{wave3}) for dS$_4$, using an ansatz $h= e^{i(x\,k_x+y\,k_y+z\,k_z)}f[t]$, is \begin{equation}
f[t]=e^{7b\,t/4}H^{(2)}_n(e^{bt}k/b)
\end{equation}
where $H^{(2)}_n=J_n+iY_n$ is the Hankel function of the second kind, $n=i\sqrt{m^2/b^2+ 31/16}$ and $k=\sqrt{k_x^2+k_y^2+k_z^2}$.  When one uses the asymptotic expansion for large order but fixed argument to order (i.e. taking the limit $b\to 0$)~\cite{gradshtein} then we can write, after factoring out a diverging constant
\begin{equation}
f[t]\sim e^{-it\sqrt{k^2+m^2}+O(b/k)}.
\end{equation}

As a final note, as we indicated earlier, we mention that (\ref{wave4}) illuminates the special role of the seemingly arbitrarily chosen conformal factor, $\Omega$.  The ground state wave function ($m=0$ solution  of (\ref{wave4})) is constant. Note however that the metric, and with that $h$ was defined in (\ref{metric6}) with a factor $\Omega^2$ extracted. Thus the true ground state wave function (provided it is normalizable) is exactly $\Omega^2$ for every choice of our parameters. 

\subsection{Mass spectrum of excitations}
In general, (\ref{wave4}) is too complicated to be transformed to the volcano potential form, familiar in 5 dimensional brane theories, and solved using standard methods for solving a one dimensional Schrodinger equation.    Consider, however that the coordinates $\sigma=u+v$ and $\delta=u-v$ are somewhat analogous to the radial and angular coordinates in a Schr\"odinger equation. Thus, one expects, that solutions dependent on the `radial coordinate', $\sigma$, only well represent  the spectrum of excitations. Among others, the gravition state, with a constant wave function, is clearly such a state.  

Indeed, using (\ref{soln}) we can see that for functions, independent of $\delta$ (\ref{wave4}) reduces to 
\begin{equation}
-Sh''+{\rm sign}(\sigma)\left(\frac{1}{2a}\tilde\lambda \Omega^{-1}-\frac{1}{4a}\tilde\Lambda_6\Omega-a \Omega S\right)h'=m^2h,
\label{eigen}
\end{equation}
where function $S$  is given in Table 1. Notice that the coefficients of (\ref{eigen}) are independent of $\delta$, just like in a Schrodinger equation with rotational invariant potential the equation for rotational invariant states is independent of the angles.  The reason for this cancellation is that the explicit $\delta$-dependence of (\ref{wave4}) comes only through multipliers $A(\delta,\sigma)$, which in turn are associated with a differentiation with respect to $\delta$.

Before we can bring (\ref{eigen}) to a Schr\"odinger form we need to introduce a new variable, $\omega$, defined by
\begin{equation}
\omega={\rm sign}(\sigma)\int_0^{|\sigma|}\frac{ds}{\sqrt{S(s)}}.
\label{defomega}
\end{equation}
The integration in (\ref{defomega}) can be analitically performed in the case of Solution I only. After the change of variables we can eliminate the first derivative from (\ref{eigen}) and obtain a Schr\"odinger equation with a volcano potential. This is done by factoring out an appropriate function $\rho(\omega)$ from $h$.  Finally, we obtain the eigenvalue equation with the volcano potential in the form 
\begin{equation}
-\partial_\omega^2h+V(\omega)h-\delta(\omega)\frac{1}{4a}[2\lambda-\Lambda_6-4 a^2+2a S'(0)]h=m^2h,
\label{sch}
\end{equation}
where
\begin{equation}
V(\omega)=-\frac{\lambda}{2}+\Omega^2\frac{(\Lambda_6-2\lambda \Omega^{-2})^2}{64 a^2S}-\frac{1}{4}a^2S\Omega^2-\frac{\Omega }{8aS}S'(\Lambda_6-2\lambda\Omega^{-2})+\frac{3S'^2}{16S}-\frac{S''}{4}.
\label{potential}
\end{equation}
We use the notation $h$ for the eigenfunction, though strictly speaking the function $h$ in (\ref{sch}) differs by a factor of $\rho$ from that used in (\ref{eigen}).

In $\Omega$ and in $S$ the variable $\sigma$ is understood to be substituted by $\sigma(\omega)$, the solution of (\ref{defomega}). The explicit form to the potential in terms of variable $\omega$ can only be given for Solution I, but using (\ref{sch}) one can qualitatively analyze the spectrum for Solution II, as well.  

The range of variable $\omega$ and the behavior of the potential at the end of the range, $\omega_0$, plays an essential role in the nature of the mass spectrum. Naturally, that range is dependent on the choice of gauge, but the variable $\omega$ is the natural coordinate.  Anyway, the mass spectrum itself is gauge invariant. 

The $m=0$ eigenfunction of (\ref{eigen}), candidate for the graviton as long as it is normalizable, is a constant.  Then the corresponding eigenfunction of (\ref{sch}) is proportional to $[\rho(\omega)]^{-1}$ and given by
\begin{eqnarray}
h_0(\omega)&=&\exp\left\{\frac{1}{8a}\int_0^\omega\left[\Omega(\Lambda_6-2\lambda \Omega^{-2})+4a^2\Omega S-2a^2 S'\right]\frac{d\omega'}{\sqrt{S}}\right\}\nonumber\\&=&\frac{w^{1/2}}{S^{1/4}}\exp\left\{\frac{1}{8a^2}\int_1^w(\Lambda_6-2\lambda w^2)\frac{dw}{wS}\right\},
\label{ground}
\end{eqnarray}
where $\Omega=1/w$, $S$, and $S'$ must be regarded as functions of $\omega$. Here $S'$ is the derivative of $S$ with respect to $a\sigma$. 
Note now that $h_0$ has no nodes, so provided it is normalizable it is the graviton state and the ground state of the eigenvalue equation.

One can make a general comment concerning the mass spectrum.  The potential depends on $\lambda$ only through the dimensionless parameter $r$ that characterizes the function $S(\sigma)$ completely. The remaining parameter $a$ just sets the scale for the excitation spectum. The value of $r$ fixes the asymptotic curvature of the six and four dimensional spaces. Introducing the parameter $r_0=-1$ for Solution I and $r_0=-50/33$ for Solution II we have
\begin{itemize}
\item
dS$_6$ and dS$_4$ for $r<r_0$
\item
M$_6$ and dS$_4$ for $r=r_0$
\item
AdS$_6$ and dS$_4$ for $0>r>r_0$
\item
AdS$_6$ and M$_4$ for $r=0$
\item
AdS$_6$ and AdS$_4$ for $r>0$
\end{itemize}

\subsection{Normalization}

The normalization condition is very simple in the Schr\"odinger type, $`\omega$-representation' .The  norm is finite if the integral over the square of the wave function converges. It is sufficient to know the wave function at  the endpoint of the range of $\omega$, $\omega_0$ (possibly infinity), as only the singular behavior at $\omega_0$ can make the solution non-normalizable.   Even when we are not able to solve the eigenvalue equation exactly we are able to determine the behavior of the solutions at $\omega_0$. 
However, the true restriction on the eigenfunctions is stronger.  We must consider only solutions that can be orthonormalized, otherwise the small oscillation cannot be expanded into modes of definite mass.  The condition of orthogonalization is that the Wronskian, $h_1 {h_2}'-h_2{h_1}'$ vanishes at $\omega=\omega_0$ \cite{mannheim}.  If the combination $h_1h'_2$ is infinite then this condition is certainly violated. This condition is particularly important in the case of inverse square potentials, which will often make their appearance below.   

Every case in which the range of $\omega$ is finite the potential energy has the endpoint behavior  $V\simeq -\frac{1}{4} (\omega_0-\omega)^{-2}$. The only acceptable solution behaves (neglecting subleading terms of the potential) as a Bessel function, $\sqrt{x}J_0(mx)$, near $x=\omega_0-\omega=0$. The Wronskian vanishes for a pair of such solutions at different values of $m$ and the solutions are orthogonal. The Wronskian for two singular solutions,  which behave like $\sqrt{x}Y_0(m_ix)$, diverges logarithmically at $x=0$ making these solutions unacceptable. Having a unique acceptable solution implies that imposing the boundary condition at $\omega=0$ introduces the usual quantization condition and the spectrum is discrete. Then, owing to the fact that in terms of the original variables, $u$ and $v$ a zero mass solution (independent of $u$ and $v$) exists, there is a physical graviton state. 

Note that the normalizability and orthogonalizability of the solutions of the Schrodinger-like equation in variable $\omega$ does not insure the finiteness of the contribution of the corresponding small oscillations to quantities like the `energy' defined by Mannheim~\cite{mannheim}. That requires the convergence of integrals over the variables $u$ and $v$ of the covariant weight, $\sqrt{g}$ and the square of the small oscillation.  This constraint  come into focus for $a<0$ solutions, for which, depending on $r$, the end of range of $w=1+a(|u|+|v|)$ may be $w=\Omega^{-1}=0$, at which point $\int d\delta \sqrt{g}\sim w^{-5}$ has a singularity. 

In tables 2. and 3. we summarized our results for both of our solutions for $a>0$.  The nature of the solutions changes as a function of $r=-\lambda/a^2$, among others the curvature of the 6 and 4 dimensional spaces, the question whether the graviton state is isolated, the order of level spacing of the spectrum, and the range of mass excitations with a weak coupling is displayed. The last two columns are important only for $\lambda<<\Lambda$, i.e. $|r|<<1$. This is the range in which corrections to the short and long range behavior of gravity do not contradict to observations in our current universe.

\begin{table}
\caption{\label{label} Solution I: Curvature of space, range of $\omega$, properies of the excitation spectrum  and the range of weakly coupled states, as a function of parameter $r$.}
\begin{center}
\begin{tabular}{llllll}
\hline
r & space & Graviton state&Level spacing& Weakly coupled\\
\hline
$r>0$& AdS$_{6}$, AdS$_{4}$ &isolated & $\Delta m\sim (-\lambda)^{1/2}$& $m<(-\Lambda)^{1/2}$\\
$r=0$& AdS$_{6}$, M$_{4}$ & marginal & continuous spectrum& $m<(-\Lambda)^{1/2}$\\
$-1<r<0$& AdS$_{6}$, dS$_{4}$ &isolated& $\Delta m\sim (-\lambda)^{1/2}$& $m<(-\Lambda)^{1/2}$\\
$r=-1$&M$_{6}$, dS$_{4}$ &  isolated & $\Delta m\sim (-\Lambda)^{1/2}$& none\\
$<=-1$&dS$_{6}$, dS$_{4}$ & isolated & $\Delta m\sim (-\Lambda)^{1/2}$& none\\
\hline
\end{tabular}
\end{center}
\end{table}

\begin{table}
\caption{\label{label} Solution II: Curvature of space, range of $\omega$, properties of the excitation spectrum  and the range of weakly coupled states, as a function of parameter $r$.}
\begin{center}
\begin{tabular}{llllll}
\hline
r & space & Graviton state&Level spacing& Weakly coupled\\
\hline
$r>r_{0}$& AdS$_{6}$, AdS$_{4}$ & isolated & $\Delta m\sim (-\Lambda)^{1/2}$& none\\
$r=r_{0}$& AdS$_{6}$, AdS$_{4}$ & isolated & continuous spectrum& none\\
$r_{0}>r>0$& AdS$_{6}$, AdS$_{4}$ & isolated & $\Delta m\sim (-\lambda)^{1/2}$& $m<(-\Lambda)^{1/2}$\\
$r=0$& AdS$_{6}$, M$_{4}$ & marginal & continuous spectrum& $m<(-\Lambda)^{1/2}$\\
$-50/33<r<0$& AdS$_{6}$, dS$_{4}$ & isolated& $\Delta m\sim (-\lambda)^{1/2}$& $m<(-\Lambda)^{1/2}$\\
$r=-50/33$&M$_{6}$, dS$_{4}$ & isolated & $\Delta m\sim (-\Lambda)^{1/2}$& none\\
$<=-50/33$&dS$_{6}$, dS$_{4}$ & isolated & $\Delta m\sim (-\Lambda)^{1/2}$& none\\
\hline
\end{tabular}
\end{center}
\end{table}

\subsection{Mass spectrum for $a<0$}
If $S(\sigma)$ has no zero in the interval $0<w<1$ (here $w=1-|a|\sigma$) then the solution is always unacceptable due to the divergence of the measure at $w=0$.   In all cases when a zero exists the parameter $r$ is large.  Though these solutions are mathematically acceptable, they could not be used for phenomenological purposes, for reasons discussed in the $a>0$ case. 
\section{Conclusions}

We solved the six dimensional Einstein equations and Israel junction
conditions in a space with mirror symmetry and exchange symmetry, $
(Z_2)^3$, in the two extra dimensions, when the 4-dimensional space
at the intersection of two 5-dimensional branes is warped.  Due to
their coordinate dependence, the junction conditions can only be
solved if the gauges in the six and five dimensional actions are
misaligned. We found two solutions, one with induced gravity terms on
the branes, and one with an induced cosmological constant only. For
each of these solutions we found two possible alternatives, having
conformal factors that are increasing and decreasing, respectively,
as the modulus of the 5th or 6th coordinate increases.

We also studied small perturbations around our solutions. Due to the
complexity of the equations satisfied by these perturbations we
restricted ourselves to those depending only on the sum of the extra
coordinates.   One of the oscillatory solutions of these equations is
the massless graviton state, which is always localized at the
intersection of the 4-branes.  In addition there is an infinite set
of solutions corresponding to gravitational excitations.   The
analytic form of the `volcano potential' could only be derived for
the solution with induced 5 dimensional gravity, but the graviton
state and the nature of the gravitational excitation spectrum could
always be analyzed.

We found that for increasing conformal factors the graviton state is
not normalizable. This is similar to the situation found in 5
dimensions by Karch and Randall~\cite{karch}. However, here the
lowest lying normalizable state is massive, leading to an
unacceptable short range gravitational force.

For decreasing conformal factor, depending on the values of the
appropriately normalized 6-dimensional and 4-dimensional cosmological
constants, $\tilde\Lambda_6$ and $\tilde\lambda$,  we obtained a
variety of spectra. For most values of the cosmological constants we
found an acceptable spectrum, with a zero mass graviton localized at
the intersection of the two branes and a gravitational excitation
spectrum that can be tuned not to lead to contradictions with
experiments verifying Newton's law in the laboratory, provided $\tilde
\Lambda_6>>\tilde\lambda$.  This work is an extension of similar
investigations in 5-dimension~\cite{karch}~\cite{friedman}\cite{xxx, yyy}
to 6 dimension. Previous studies of a 6 dimensional model with
similar symmetries allowed only for  M$_4$ space at the intersection
of the branes~\cite{arkani}.

The model we studied requires the input of the value of the 4-
dimensional cosmological constant by hand.  We find this esthetically
not very satisfactory.  This blemish could in principle be
circumvented by the addition of a global scalar, or several scalars,
like quintessence, which could possibly generate a cosmological
constant which may even depend on time.  We intend to return to this
problem in a future publication.

We are somewhat perplexed by the existence of two acceptable
solutions at given values of the input parameters (cosmological
constants). We have not been able to find a physical principle which
would allow us to discard one of the solutions yet.  One possibility
is to investigate observable post-Newtonian effects.
The models under discussion have two separate gravity actions in 6 and 5 dimensions. Therefore, just like the Dvali-Gabadadze-Porrati (DGP) model~\cite{DGP}, they could suffer from the van Dam-Veltman-Zakharov ~\cite{vDVZ, zachar} type of discontinuity and could be severely constrained. For more details we refer the reader to the following papers~\cite{ABC,daffayet,porrati,tanaka}. 

The research  reported here  was  supported in part  by the  U.S. Department of Energy  grant  No  DE-FG02-84ER40153. 

\appendix
\setcounter{section}{0}
\setcounter{equation}{0}
\section{Solution of the global equations}

The two Einstein equations that we use to find the solution are then
\begin{eqnarray}
{\cal G}_{uu}&=&g_{uu}\left[\frac{10 a^2}{g(u,v)}+\frac{10 a^2}{g(v,u)}-2\tilde\lambda(1+a(u+v))^2\right.\nonumber\\&-&\left.(1+a(u+v))\frac{2a}{g(v,u)}\left(\frac{g_{,u}(v,u)}{g(u,v)}-\frac{g_{,v}(v,u)}{g(v,u)}\right)\right]=-\tilde\Lambda_6g_{uu},\label{one}\\
{\cal G}_{uv}&=&-\frac{2 a}{1+a(u+v)}\left(\frac{g_{,v}(u,v)}{g(u,v)}+\frac{g_{,u}(v,u)}{g(v,u)}\right)=0,
\label{two}
\end{eqnarray}
where $\tilde\lambda=8\pi G_4\lambda$.
Substituting $\frac{g_{,u}(v,u)}{g(v,u)}$ from (\ref{two}) into (\ref{one}) we obtain a first order differential equation for the symmetric function
\begin{equation}
F(u,v)=\frac{1}{2g(u,v)}+\frac{1}{2g(v,u)}
\label{fdef}
\end{equation}
\begin{equation}
20a^2F(u,v)-2\lambda(1+a(u+v))^2+\tilde\Lambda_6-(1+a(u+v))4aF_{,v}(u,v)=0
\end{equation}
First of all, this equations implies that the derivative $F_{,v}(u,v)$ is also symmetric. That is possible only if $F(u,v)$ is a function of $\sigma=u+v$ only.  Keeping this in mind it has an analytic solution
\begin{equation}
F(u,v)\equiv S(\sigma)=\frac{1}{a^2}\left[(1+a\sigma)^2\frac{\lambda}{6}-\frac{\tilde\Lambda_6}{20}+C (1+a\sigma)^5\right],
\end{equation}
where $C$ is an integration constant.

Now defining $\delta=u-v$ and using (\ref{fdef}) we can write $g(u,v)$ in terms of an odd function of $\delta$, $A(\delta,\sigma)$ as
\begin{equation}
g(u,v)=\frac{1}{S(\sigma)+A(\delta,\sigma)}
\label{gform}
\end{equation}

Inserting (\ref{gform}) into (\ref{two}) we obtain a differential equation for $A(\delta,\sigma)$
\begin{equation}
A(\delta,\sigma)A_{,\sigma}(\delta,\sigma)+S(\sigma)[A_{,\delta}(\delta,\sigma)-S'(\sigma)]=0,
\label{aeq}
\end{equation} 
We have not been able to give a solution of (\ref{aeq}) in closed form, but the terms of its Taylor series in the variable $\delta$ can be readily calculated. We will write
\begin{equation}
A(\delta,\sigma)=\sum_{k=0}^\infty \delta^{2k+1}\alpha_k(\sigma).
\label{series}\end{equation}
Then by substituting (\ref{series}) into (\ref{aeq}) we obtain
\begin{equation}
\alpha_0(\sigma)=S'(\sigma).
\label{alpha0}
\end{equation}
and the recursion relation
\begin{equation}
\alpha_k(\sigma)=-\frac{1}{(2k+1)S(\sigma)}\sum_{m=0}^{k-1}\alpha_{k-m-1}(\sigma)\alpha_m'(\sigma).
\label{recursion}
\end{equation}
Using the recursion relation we can easily generate the coefficients $\alpha_k(\sigma)$ in arbitrary order. We obtain
\begin{eqnarray*}
\alpha_1(\sigma)&=&-\frac{ S'(\sigma )\,
      S''(\sigma ) }
    {3\,S(\sigma )}.\\
\alpha_2(\sigma)&=&\frac{S'(\sigma )\,
    \left( -{S'(\sigma
          )}^2\,S''(\sigma )
         + 
      2\,S(\sigma )\,
       {S''(\sigma )}^2 + 
      S(\sigma )\,
       S'(\sigma )\,
       S^{(3)}(\sigma )
      \right) }{15\,
    {S(\sigma )}^3},\\
\alpha_3(\sigma)&=-&\frac{S'(\sigma )}   {315\,
    {S(\sigma )}^5}\,
      \left[ 9\,
         {S'(\sigma )}^4\,
         S''(\sigma ) + 
        17\,{S(\sigma )}^2\,
         {S''(\sigma )}^3 - 
        9\,S(\sigma )\,
         {S'(\sigma )}^3\,
         S^{(3)}(\sigma ) \right.\\& + &\left. 
        26\,{S(\sigma )}^2\,
         S'(\sigma )\,
         S''(\sigma )\,
         S^{(3)}(\sigma )+
        S(\sigma )\,
         {S'(\sigma )}^2\,
         \left( -29\,
           {S''(\sigma )}^2 + 
           3\,S(\sigma )\,
           S^{(4)}(\sigma )
           \right)  \right]
  ,\\
...
\end{eqnarray*}

For the special case of $\Lambda_6=0$, i.e. assuming that the six dimensional cosmological constant vanishes, still we get a curved space on the 3+1 dimensional word.  In fact, at least in solution 1,   $A(\delta, \sigma)$ simplifies considerably. We will consider this case next.  

First of all, $S(\sigma)=w^2$, where we introduced the notation $w=1+a \sigma$.  If we also introduce the notation 
\begin{equation}
z=1-\frac{a\delta}{1+a\sigma}
\end{equation}
and we regard A as a function of $w$ and $z$, then  $A$ can be written in the  form
\begin{equation}
A=w^2h(z),
\end{equation}
where $h(z)$ satisfies the ordinary differential equation
\begin{equation}
2h^2+hh'(1-z) -h'=2
\label{hdiff}
\end{equation}
We know the expansion of $h(z)$ around $z=1$ already.  It is quite interesting to investigate $h(z)$ around $z=0$, as well. The exact point, $z=0$ is not physical as $a\delta<1+a\sigma$, but at large $u$ and small $v$ (near the $u$ axis) $z$ approaches 0.  Numerical solution of this equations shows that $h(0)$=1 and $ h'(1)=0.$ Of course we know that $h(1)$=0.  Then we can write $h(z)=1-k(z)$ and expand (\ref{hdiff}) in $k$, keeping linear terms only. Simple calculation shows that (\ref{hdiff}) reduces to $-4 k+z k'=0$ leading to $h(z)=1-c z^4+O(z^7)$ (here we also calculated the next to leading term of $k$).  This form is consistent with the expansion of (\ref{hdiff}) in a power series of $k$. The constant $c$ can be identified from the numerical calculation as $c=1/4$.

The case $\Lambda_6\not=0$ can also be investigated analytically near $z=0$ (near the axes, far from the origin). Using the ansatz
\begin{equation}
A(\delta,\sigma)= S(w)-h(z)f(w)
\end{equation}
(\ref{aeq}) can be expanded to retain the first power of $h$ only. One obtains a separable equation
\begin{equation}
h(z)[S(w)f'(w)+f(w)S'(w)]-S(w)f(w)h'(z)z=0
\end{equation}
The solution of this equation is
\begin{eqnarray}
f(w)&=&\frac{w^n}{S(w)},\nonumber\\
h(z)&=&c z^n.
\label{asymp}
\end{eqnarray}
Note that in the special case $S=w^2$ the integration constant $n=4$. 

\section{Junction conditions with gauge transformations}

When we perform (\ref{transform}) the components of the metric tensor, $g_{uu}$, $g_{vv}$ and $g_{uv}$ change in the following way (the time-like and ordinary space-like components are unchanged aside from the replacements of the variables )
\begin{eqnarray}
g_{uu}&=&\frac{g(u,v)}{1+a(u+v))^2}\to\tilde  g_{uu}=R\left[\tilde g(u,v)[h_{,v}(v,u)]^2+\tilde g(v,u)[h_{,u}(v,u)]^2\right],\nonumber\\
g_{vv}&=&\frac{g(v,u)}{(1+a(u+v))^2}\to\tilde g_{vv}=R\left[\tilde g(v,u)[h_{,u}(u,v)]^2+\tilde g(u,v)[h_{,v}(u,v)]^2\right],\nonumber\\
g_{uv}&=&0\to\tilde g_{uv}=-R\left[ \tilde g(v,u)h_{,u}(u,v)h_{,u}(u,v)+\tilde g(u,v)h_{,v}(v,u)h_{,v}(u,v)\right],
\label{metrictransform}
\end{eqnarray}
where we use the notation
$\tilde g(u,v)=g(h(u,v),h(v,u)) $, $\tilde g(v,u)=  g(h(v,u),h(u,v))$ and where
\begin{equation}
R=\frac{1}{[h_{,u}(u,v)h_{,v}(v,u)-h_{,v}(u,v)h_{,u}(v,u)]^{2}\{1+a[h(u,v)+h(v,u)]\}^2}.
\end{equation}

Now notice that the components of the gauge transformed metric tensor at $u=v=0$ are given by
$\tilde g_{uu}(0,0)=\tilde g_{vv}(0,0)=g(0,0)[h_{,u}(u,v)|_{u=v=0}]^2.$ Since normalization requires that these quantities are equal to 1 we have imposed the condition  $h_{,u}(u,v)|_{u=v=0}=1$ on (\ref{transform}).

In this Appendix we will prove that the gauge transformed junction contributions of the six dimensional theory can be cancelled by the five dimensional curvature scalar and cosmological constant terms.   First we need to find the gauge transformed form of the junction contributions (\ref{junction6}).  We will concentrate only on the contribution at the $v=0$ brane as once we solved the junction conditions at the $v=0$ brane the junction conditions at the $u=0$ brane are automatically satisfied considering the symmetry of gauge transformations (\ref{transform}).  As a reminder, we must take linear combination of the $\Delta \tilde R^{(6)}_{uu}$,  $\Delta \tilde R^{(6)}_{vv}$ and  $\Delta \tilde R^{(6)}_{uv}$ contributions, such that they correspond to functional derivatives with respect to the untransformed metric components. These combinations are
\begin{eqnarray}
\Delta_v R^{(6)}_{\mu\mu}&=&-g_{\mu\mu}\frac{1+au)^2}{1+ah}\delta(v)\left[h_{,v}(v,u)|_{v=0}\right]^3\frac{8\,a\,g(h,0)-(1+a\,h)g_{,v}(h,v)|_{v=0}}{g(h,0)g(0,h)},\nonumber\\
\Delta_v R^{(6)}_{uu}&=&-g_{uu}\frac{(1+au)^2}{1+ah}\delta(v)[h_{,v}(v,u)|_{v=0}]^3\frac{8ag(h,0)}{g(0,h)},
\label{newjunction}
\end{eqnarray}
where we used the abbreviated notation $h=h(u,0)$.

Now we are ready to substitute (\ref{vzerobrane}) and (\ref{newjunction}) into junction condition (\ref{junction3}). Note that the two conditions ($J_1(u)=0$ and $J_2(u)=0$) at $AB=\mu\mu$ and at $AB=uu$ contain two independent functions: $k(u)=h(u,0)$ and $f(u)=h_{,v}(v,u)|_{v=0}$.  These two functions are unconstrained aside from the following conditions: $k(0)=0$, $k'(0)=1$, $f(0)=1$. $f(u)$ appears in a very simple manner in the two junction conditions as shown by (\ref{newjunction}). We can easily eliminate this function from the two junction conditions by taking the ratio of the two junction conditions. The resulting condition (call it $F(k(u),u)=0$) is a transcendental function of $k(u)$  (not a differential equation) and, in principle, can be solved for it. It has the form
\begin{equation}
\frac{8\,a\,g(k,0)-(1+a\,k)g_{,v}(k,v)|_{v=0}}{8ag(k,0)}=\frac{\lambda(1+au)^2-\tilde\Lambda_5-\frac{6a^2}{g(u,0)}-\frac{3ag_{,u}(u,0)(1+au)}{2 [g[u,0)]^2}}{2\lambda (1+au)^2-8\pi G_5\Lambda_5-\frac{6a^2}{g(u,0)}}
\end{equation}
 Once we solved for $k(u)$ we can calculate  $f(u)$ as well.  All this can be done provided the three constraints on $k(u)$ and $f(u)$ are satisfied. Note that due to these three constraints the junction conditions $J_1(0)=0$, $J_2(0)=0$, and $dF(h(u),u)/du|_{u=0}=0$ become independent of the form of the gauge transformations.  All other (higher) derivatives of the junction conditions contain unconstrained derivatives of the gauge transformation function. These provide equations for those unconstrained derivatives.  We can conclude that the junction conditions can be solved for the gauge transformations provided three constraints among the constants of the theory are satisfied. The conditions $J_1(0)=0$ and $J_2(0)=0$  were given earlier in (\ref{uno}) and (\ref{due}), while the condition $dF(f_1(u),u)/du|_{u=0}=0$ was given in (\ref{third}).

\end{document}